\documentclass[%
 reprint,
superscriptaddress,
nofootinbib,
nobibnotes,
 amsmath,amssymb,
prb,
]{revtex4-1}
\usepackage{graphicx}
\usepackage{dcolumn}
\usepackage{bm}
\usepackage{notoccite}
\usepackage{physics}
\usepackage{xcolor}
\usepackage{mathtools}
\begin{document}

\title{Spin field-effect transistor in a quantum spin-Hall device}

\author{Raffaele Battilomo}
\affiliation{Institute for Theoretical Physics, Center for Extreme Matter and Emergent Phenomena, Utrecht University, Princetonplein 5, 3584 CC Utrecht,  Netherlands}
\author{Niccol\'o Scopigno}
\affiliation{Institute for Theoretical Physics, Center for Extreme Matter and Emergent Phenomena, Utrecht University, Princetonplein 5, 3584 CC Utrecht,  Netherlands}
\author{Carmine Ortix}
\affiliation{Institute for Theoretical Physics, Center for Extreme Matter and Emergent Phenomena, Utrecht University, Princetonplein 5, 3584 CC Utrecht,  Netherlands}
\affiliation{Dipartimento di Fisica ``E. R. Caianiello", Universit\'a di Salerno, IT-84084 Fisciano, Italy}

\begin{abstract}
We discuss the transport properties of a quantum spin-Hall insulator with sizable Rashba spin-orbit coupling in a disk geometry. The presence of topologically protected helical edge states allows for the control and manipulation of spin polarized currents: when ferromagnetic leads are coupled to the quantum spin-Hall device, the ballistic conductance is modulated by the Rashba strength. Therefore, by tuning the Rashba 
interaction via an all-electric gating, it is possible to control the spin polarization of injected electrons.
\end{abstract}

\maketitle

\section{Introduction}
\label{sec: intro}
Spintronics \citep{Spintronics} is the field dedicated to studying how to actively control and manipulate the electronic spin degree of freedom in solid-state systems. When spin-orbit coupling (SOC) is present, the electron's spin and momentum are locked to each other allowing us to study the interplay between charge and spin
degrees of freedom. This interplay is of central interest since it opens the possibility to manipulate
electric currents by controlling the electronic spin and viceversa, paving the way to the development of new devices of technological relevance.
The first proposal of a spintronic device with electrical spin manipulation was the spin field-effect transistor, brought forward by Datta and Das\citep{dat1990}. Since then, spin transistors have been the subject of intense research and, to date, are still a central problem in the field of spintronics. In recent years, much attention has been drawn to exploit spin interference effects on the electronic transport properties of mesoscopic devices with loop geometries, such as semiconducting quantum rings\citep{Nitta1999,Naga2013,Fru2004,Molnar2004,Kon2006,Saarikoski2018,Ying2016}.  
In the presence of electromagnetic potentials, the conductance of a semiconductor ring exhibits signs of quantum interference due to the Aharonov-Bohm\citep{Bohm1959} (A-B) and Aharonov-Casher\citep{AC1984}(A-C) effects. 
Being manifestations of the Berry phase, these interference effects have been exploited to detect in the laboratory the  spin geometric phase \cite{Naga2013,Naga2012}.

An alternative setup for the detection of the $\pi$-Berry phase was proposed in Ref. \onlinecite{Chen2016}, with the interferometer based on a quantum spin-Hall (QSH) insulator. The QSH insulator\citep{Ber2006} is a time-reversal symmetric topological state of matter which possesses in gap helical edge states: at each edge of the system there are two 
counterpropagating states with opposite spin projections. The realization of QSH insulators in HgTe\citep{Kon2007} and InAs/GaSb/AlSb\citep{Liu2008} quantum wells has since opened the possibility to engineer new types of spin transistors. A notable example has been discussed in Ref.~\onlinecite{Mac2010} where by combining the helical nature of the QSH edge states with the A-B effect, a setup that behaves as a spin transistor
has been proposed. One of the main advantages of using a QSH insulator lies in the fact that transport in the system is ballistic in nature and takes place along the edges, making it effectively a one-dimensional (1D) system. 

In this paper we consider an alternative QSH-based setup exploiting \textcolor{black}{the effect of the Rashba spin-orbit coupling (SOC), originating from structural inversion asymmetry}. \textcolor{black}{The presence of Rashba SOC\citep{Rot2010} breaks the axial spin symmetry of the helical edge states, tilting their spin projection in the QSH plane. Consequently, in a disk geometry the Rashba SOC yields a local rotation of the spin projection of the helical edge states along the disk}. By \textcolor{black}{further} controlling the Rashba strength through an additional external gate voltage it is possible to both modulate electric currents which pass through the QSH insulator and manipulate the spin projection of single incoming electrons, similarly to the A-B based setup of Ref.~\onlinecite{Mac2010}. 
By studying the transport properties of the QSH insulator in the presence of Rashba SOC we will indeed show how such a system can be used as an all-electric spin transistor. It is interesting to note that since there are only two counterpropagating modes at the edge, the QSH insulator can be seen as a faithful implementation of the original Datta-Das transistor: the original device introduced in Ref.~\onlinecite{dat1990} drew inspiration from an electro-optic modulator in which polarized light was split into two beams that suffered different phase shifts. In this sense the QSH is a true electronic analog of the electro-optic modulator.\\
\indent The paper is organized as follows: in Section \ref{sec:BHZ} we study the Bernevig-Hughes-Zhang\citep{Ber2006} (BHZ) Hamiltonian with the addition of a linear Rashba SOC term. We find the in gap eigenstates and eigenvalues of the system in a disk geometry and show how the Rashba coupling is responsible for the tilting of the spin projection of the two helical edge states. In Section \ref{sec:1D} we study an effective 1D model of the QSH disk and calculate the conductance of the system through the Landauer approach. We show that the QSH disk behaves as a spin field-effect transistor. Finally in Section \ref{sec:kwant}, by using the microscopic tight-binding BHZ model, we validate our findings through a numerical calculation of the conductance.
\section{ In gap states of the BHZ Hamiltonian in the Presence of Rashba spin-orbit coupling}
\label{sec:BHZ}
We begin by studying the properties of the helical edge states of a QSH insulator disk in the presence of Rashba SOC~\cite{Michetti2011,Ortiz2016,Rod2015}. 
It is well known\citep{Kon2007} that the QSH phase occurs in the ``inverted'' regime of HgTe/CdTe semiconductor quantum wells, which is achieved by tuning the thickness of the HgTe well above a critical thickness $d_c \simeq 6~$nm. The occurrence of this topological phase transition can be captured using conventional $\mathbf{k}\cdot\mathbf{p}$ theory. Starting from the six-band Kane model\cite{Winkler} and using  perturbation theory near the $\Gamma$ point one can obtain an effective four band model ($\left |E1,j_z=\pm \frac{1}{2}\right\rangle$, $\left |H1,j_z=\pm \frac{3}{2}\right\rangle$) for the subbands of the quantum well structure \citep{Ber2006}. The $\left |E1,j_z=\pm \frac{1}{2}\right\rangle$ subbands are a linear combination of s-like $\left |\Gamma_6,j_z=\pm \frac{1}{2}\right\rangle$ and the light hole $\left |\Gamma_8,j_z=\pm \frac{1}{2}\right\rangle$ bands while the $\left |H1,j_z=\pm \frac{3}{2}\right\rangle$ come about from the $\left |\Gamma_8,j_z=\pm \frac{3}{2}\right\rangle$ heavy hole bands. The electronic structure is then described by the effective BHZ Hamiltonian,
\begin{equation}
\mathcal{H}_{BHZ}=
\begin{psmallmatrix}
\epsilon_\mathbf{k} +M_\mathbf{k}& Ak_+ & 0 & 0 \\
Ak_- &\epsilon_\mathbf{k} -M_\mathbf{k}  & 0 & 0 \\
0 & 0 & \epsilon_\mathbf{k} +M_\mathbf{k}& -Ak_-\\
0 & 0 & -A k_+ & \epsilon_\mathbf{k} -M_\mathbf{k}
\end{psmallmatrix}.
\label{Hbhz}
\end{equation}
Eq.~\ref{Hbhz} is written in the basis $\left |E1, j_z= +\frac{1}{2}\right\rangle$,  $\left |H1, j_z= +\frac{3}{2}\right\rangle$, $\left |E1, j_z=- \frac{1}{2}\right\rangle$ and $\left |H1,j_z=- \frac{3}{2}\right\rangle$. We have defined $\epsilon_\mathbf{k}=C-D(k_x^2+k_y^2)$, $M_\mathbf{k}=M-B(k_x^2+k_y^2)$ and $A,\, B,\, C,\,D$ and $M$ are model parameters. We have also introduced $k_\pm = k_x \pm i k_y$ with $k_{x,y}=-i \partial_{x,y}$. For simplicity, we set $C=D=0$. The Hamiltonian above preserves time-reversal symmetry, with the time reversal symmetry operator defined as $\Theta=-i (\sigma_y\otimes \sigma_0) \mathcal{K}$, where $\mathcal{K}$ stands for complex conjugation and $\sigma_\alpha$ are the Pauli matrices. When $sign(B)=sign(M)$ the system is in the QSH phase and is characterized by a nontrivial ${\mathbb Z}_2$ topological invariant.

\begin{figure}[tbp]
\includegraphics[width=\columnwidth]{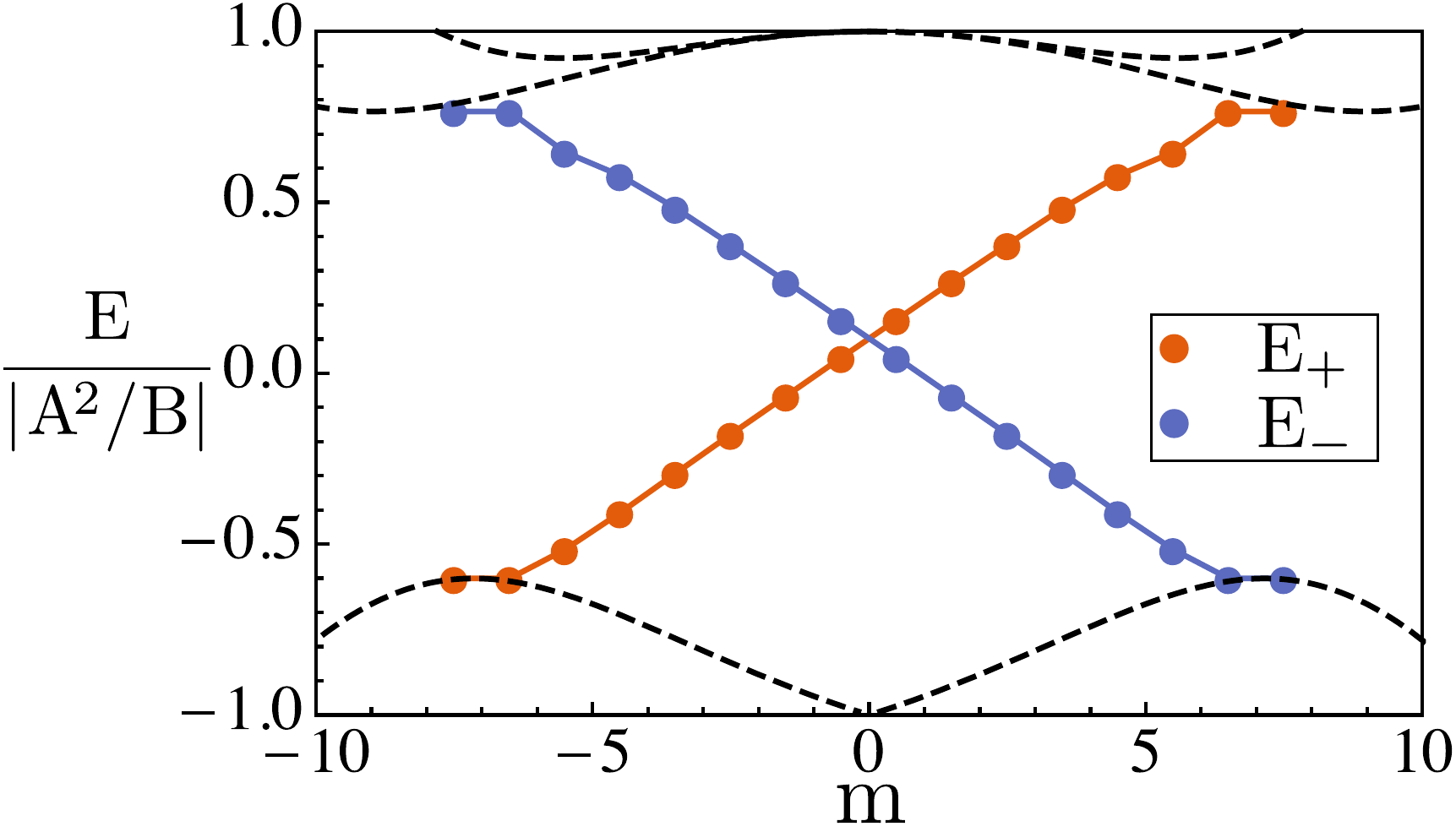}
\caption{Energy dispersion of the two in-gap helical edge states of a QSH insulator in a disk geometry \textcolor{black}{as a function of the half-integer eigenvalue m of the operator $\Sigma_z$}. $E_\pm$ indicate respectively clockwise movers and counterclockwise movers. The dashed lines represent the bulk bands. The value of the Rashba strength is set to $\alpha_R = |A|/2$. The maximum value of the Rashba strength compatible with the presence of a full bulk band gap is $\alpha_R^{max}=2 |A|$. Energies have been measured in units of $|A^2/B|$.}
\label{edgestates}
\end{figure}

The presence of structural inversion asymmetry gives rise to  Rashba terms\citep{Rot2010} which couple the two spin blocks in Eq.~\ref{Hbhz} and break the axial spin symmetry. At linear order in $\mathbf{k}$ the Rashba Hamiltonian only couples the $\left |E1,j_z= \pm\frac{1}{2}\right\rangle$ bands,
\begin{equation*}
\mathcal{H}_{R}=
\begin{pmatrix}
0 & 0 &- i \alpha_R k_- &0\\
0 &0 &0 &0\\
i\alpha_R k_+ &0 &0 &0\\
0 &0 &0 &0
\end{pmatrix}.
\end{equation*}
\textcolor{black}{It is easy to see that unlike $\mathcal{H}_{BHZ}$, $\mathcal{H}_{R}$ breaks the effective two-dimensional inversion symmetry, $\mathcal{H}_R(k)\neq \mathcal{I}_{2D}\,\mathcal{H}_R(-k)\,\mathcal{I}_{2D}$, where $\mathcal{I}_{2D}=\sigma_z \otimes \sigma_0$ is the inversion operator.}\\
\indent To find the helical edge states dispersion of the full Hamiltonian  $\mathcal{H}=\mathcal{H}_{BHZ}+\mathcal{H}_R$ in a disk geometry, we write $\mathcal{H}$ in polar coordinates $(r,\phi)$:
\begin{equation}
\mathcal{H}=
\begin{psmallmatrix}
B \, \Pi+M &-iA\, \Lambda^+  & -\alpha_R\, \Lambda^- &0 \\
-iA\, \Lambda^-&-B\,\Pi-M& 0 &0\\
\alpha_R\,\Lambda^+& 0 &B\,\Pi+M & iA\,\Lambda^-\\
0 & 0 & iA\,\Lambda^+  & -B\,\Pi-M
\end{psmallmatrix},
\label{polarH}
\end{equation}

where
\begin{align*}
&\Pi=(\partial_r^2+\frac{1}{r}\partial_r+\frac{1}{r^2}\partial_\phi^2),\\
&\Lambda^+=e^{i\phi}(\partial_r+\frac{i}{r}\partial_\phi),\\
&\Lambda^-=e^{-i\phi}(\partial_r-\frac{i}{r}\partial_\phi).
\end{align*}

We are now interested in solving the Schr\"{o}dinger equation $\mathcal{H}\psi (r,\phi)=E \psi (r,\phi)$ for energies inside the insulating bulk gap. In order to diagonalize the Hamiltonian of Eq.~\ref{polarH} we first show that the problem is separable in the two variables $r$ and $\phi$. The total electronic angular momentum is given by the composition of the spin angular momentum with the \textcolor{black}{orbital angular momentum}, $\mathbf{J}=\mathbf{L}+\mathbf{S}$, plus the angular momentum $\mathbf{L_ {\boldsymbol{\phi}}}$ due to the rotation around the disk. Its projection along the $\hat{z}$ axis is given by the operator $\Sigma_z=L_\phi+J_z$,  with $L_\phi=-i\partial_\phi$ and $J_z=diag[\frac{1}{2},\frac{3}{2},-\frac{1}{2},-\frac{3}{2}]$. It is straightforward to show that $[\mathcal{H},\Sigma_z]=0$,  and hence our wave functions can be written as eigenvectors of $\Sigma_z$. The in gap solutions for Eq.~\ref{polarH} take the form,
\begin{equation*}
\psi^\xi_m(r,\phi)=
\begin{pmatrix}
e^{i(m-\frac{1}{2})\phi}c_1(\xi)I_{m-\frac{1}{2}}(\xi r)\\
e^{i(m-\frac{3}{2})\phi}c_2(\xi)I_{m-\frac{3}{2}}(\xi r)\\
e^{i(m+\frac{1}{2})\phi}c_3(\xi)I_{m+\frac{1}{2}}(\xi r)\\
e^{i(m+\frac{3}{2})\phi}c_4(\xi)I_{m+\frac{3}{2}}(\xi r)
\end{pmatrix}
\end{equation*}
where $m$ is the half-integer eigenvalue of the operator $\Sigma_z$ ensuring the $2 \pi$ periodicity of the wave functions, whereas
$I_m(r)$ are the modified Bessel functions of the first kind, necessary to have a normalizable solution. Finally, $c$ and $\xi$ are constants which depend on the system's parameters and energy. \textcolor{black}{For a given in gap energy $E$ we find four values of $\xi$ for which the wave functions $\psi_m^{\xi}$ are linearly independent}: the total wave function can then be written as a linear combination,
\begin{equation}
\Phi_m(r,\phi)=\sum_{i=1}^4 a_i \,\psi^{\xi_i}_m(r,\phi).
\end{equation}
To find the in gap eigenvalues we impose fixed boundary conditions at the edge of the disk: $\Phi_m(r=r_0,\phi)=0$. Fig. \ref{edgestates} shows the energy dispersion of the helical edge states in the QSH phase. Away from the energy bulk the dispersion is practically linear as conventionally found in ribbon geometries\citep{Ortiz2016,Zhang2012}.

\begin{figure}[tbp]
\includegraphics[width=\columnwidth]{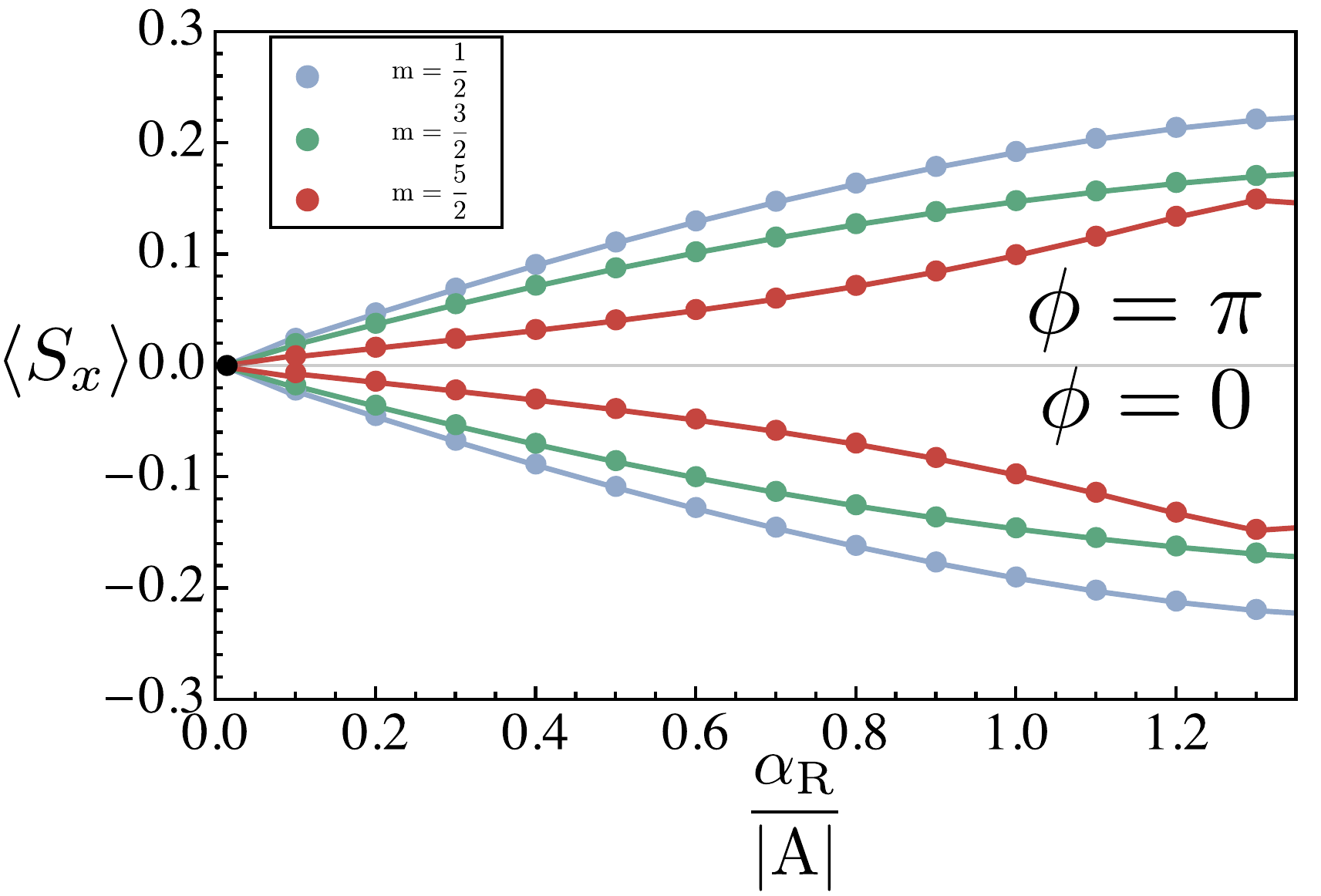}
\caption{Local expectation value of $S_x$ for different counterclockwise moving states as a function of the Rashba strength $\alpha_R$. \textcolor{black}{ The spin projection is measured at two opposite points of the disk, $\phi=0$ and $\phi=\pi$, where the only nonzero in-plane spin component is given by $S_x$. The values of $\langle S_x \rangle$ have been computed numerically and are represented by dots in the above graph; lines joining the dots are present only as a guide for the eye.}}
\label{spintilt}
\end{figure}

\textcolor{black}{The presence of the Rashba SOC breaks the axial spin symmetry of the BHZ Hamiltonian and tilts the electronic spin towards the QSH plane. This can be seen by computing the out-of-plane spin component $S_z=\frac{1}{2}\sigma_0\otimes\sigma_z$: the local expectation value $\langle S_z \rangle$ decreases monotonically by increasing the Rashba strength. Moreover, in the current disk geometry, the in-plane spin component is reversed under a $\pi$ rotation. For instance, at $\phi=0$ and $\phi=\pi$ the only in-plane component corresponds to $S_x=\frac{1}{2}\sigma_0\otimes\sigma_x$, and as shown in Fig. \ref{spintilt} it is completely reversed after half a turn.}\\
\indent We are now interested in using this spin-tilt effect to study the electronic transport through the disk when the Fermi energy is in the bulk band gap. In this case the QSH behaves effectively as a 1D single mode ballistic conductor with only two counterpropagating states at the edge. For this reason it is inherently different from a quasi-1D semiconductor ring \citep{Fru2004}. In the latter, for all Fermi energies, there are two clockwise movers and two counterclockwise movers (neglecting transverse modes). As we will show, the difference in number of propagating channels between the two systems will  result in distinct transport properties.

 A qualitative understanding of electronic transport through the QSH can be achieved by considering an effective 1D model that retains the helical nature of the edge states and the spin-tilting mechanism of the Rashba SOC.

\section{Effective 1D Model}

\label{sec:1D}

\subsection{1D Hamiltonian}
\label{sec:effH}
In order to have two counterpropagating helical modes which mimic the QSH edge states and a Rashba spin-tilting mechanism we study the following effective 1D Hamiltonian,
\begin{equation}
\mathcal{H}_{eff}=-\frac{i \hbar}{2}\mathcal{f}\omega_z\sigma_z+\omega_R\sigma_r\,,  \partial_\phi\mathcal{g}
\label{H1D}
\end{equation}
where $\mathcal{f}\,,\mathcal{g}$ is the anticommutator, necessary to have a hermitian Hamiltonian\cite{Mei2002,Ortix2015}. In Eq. \ref{H1D} we have defined the two characteristic frequencies $\omega_z=\frac{v_F}{r_0}$ (with $v_F$ Fermi velocity of the edge states and $r_0$ disk radius) and $\omega_R=\frac{2\alpha_R}{\hbar r_0}$. The radial Pauli matrix is defined as $\sigma_r=\cos\phi\sigma_x+\sin\phi\sigma_y$.  
The eigenvalues of  Eq.~\ref{H1D} are,
\begin{equation}
E_{\pm}(m)=\hbar \omega_z \left[ -\frac{1}{2}\pm m\sqrt{1+\left(\frac{\omega_R}{\omega_z}\right)^2}\, \right].
\label{Eigenvalues1D}
\end{equation}
The spectrum, much like the one in Fig.~\ref{edgestates},  is linear \textcolor{black}{in $m$} and obeys time-reversal symmetry: at each energy $E$ there are two corresponding eigenstates with opposite spin projections and opposite velocities. The eigenstates of Eq.~\ref{H1D} are,
\begin{align}
\psi_m^+(\phi)=
e^{im\phi}
\begin{pmatrix}
e^{-i\frac{\phi}{2}}\cos\frac{\gamma}{2}\\
e^{i\frac{\phi}{2}}\sin\frac{\gamma}{2}
\end{pmatrix}
\label{eig1}
\\
\psi_m^-(\phi)=
e^{im \phi}
\begin{pmatrix}
-\,e^{-i\frac{\phi}{2}}\sin\frac{\gamma}{2}\\
e^{i\frac{\phi}{2}}\cos\frac{\gamma}{2}
\end{pmatrix}
\label{eig2}
\end{align}

where $m$ is the half-integer eigenvalue of the operator $\Sigma_z^{\textcolor{black}{eff}}=-i \partial_\phi+\frac{\sigma_z}{2}$. The angle $\gamma=\arctan(\frac{\omega_R}{\omega_z})$ measures the spin tilt with respect to the quantization axis $\hat{z}$: at zero Rashba ($\gamma=0$) the spinors in Eqs.~\ref{eig1} and \ref{eig2} simply reduce to $\left | \uparrow,\downarrow\right\rangle$ eigenstates of $\sigma_z$. This is in agreement with the zero Rashba behavior of the BHZ Hamiltonian of Eq. \ref{Hbhz}, where the edge states are eigenstates of $S_z=\frac{1}{2}\sigma_0\otimes\sigma_z$. In the following we will take into account eigenstates which lie far from the bulk bands, close to zero energy, where the dispersion is mostly linear. For these states the spin tilting effect is larger, as shown in Fig. \ref{spintilt}. Under these assumptions we can safely
describe the BHZ helical edge states with the effective 1D Hamiltonian of Eq. \ref{H1D}.
\subsection{Scattering Matrix Approach}
\label{sec:smatrix}
\begin{figure}[tbp]
\includegraphics[width=\columnwidth]{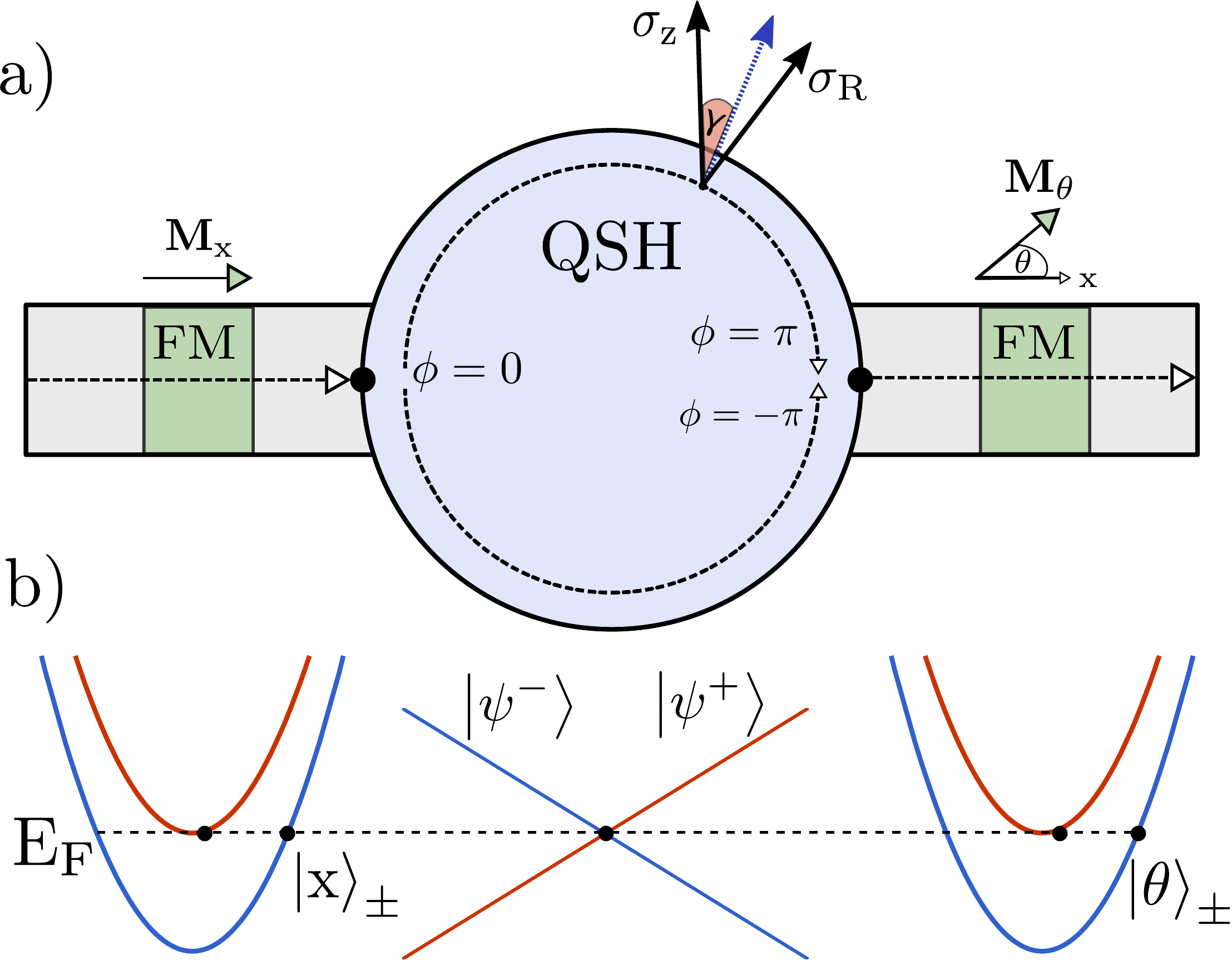}
\caption{a) QSH spin field effect transistor setup. The two semi-infinite ferromagnetic leads are coupled symmetrically at $\phi=0$ and $\phi=\pi$. Magnetizations in the left and right leads are given by the direction of $M$. Inside the disk, the electron spin is tilted by an angle $\gamma$ in the direction of the Rashba field. (b) Schematic energy dispersion of $\mathcal{H}_{FM}$ and $\mathcal{H}_{eff}$. }
\label{setup}
\end{figure}
\begin{figure}[tbp]
\includegraphics[width=\columnwidth]{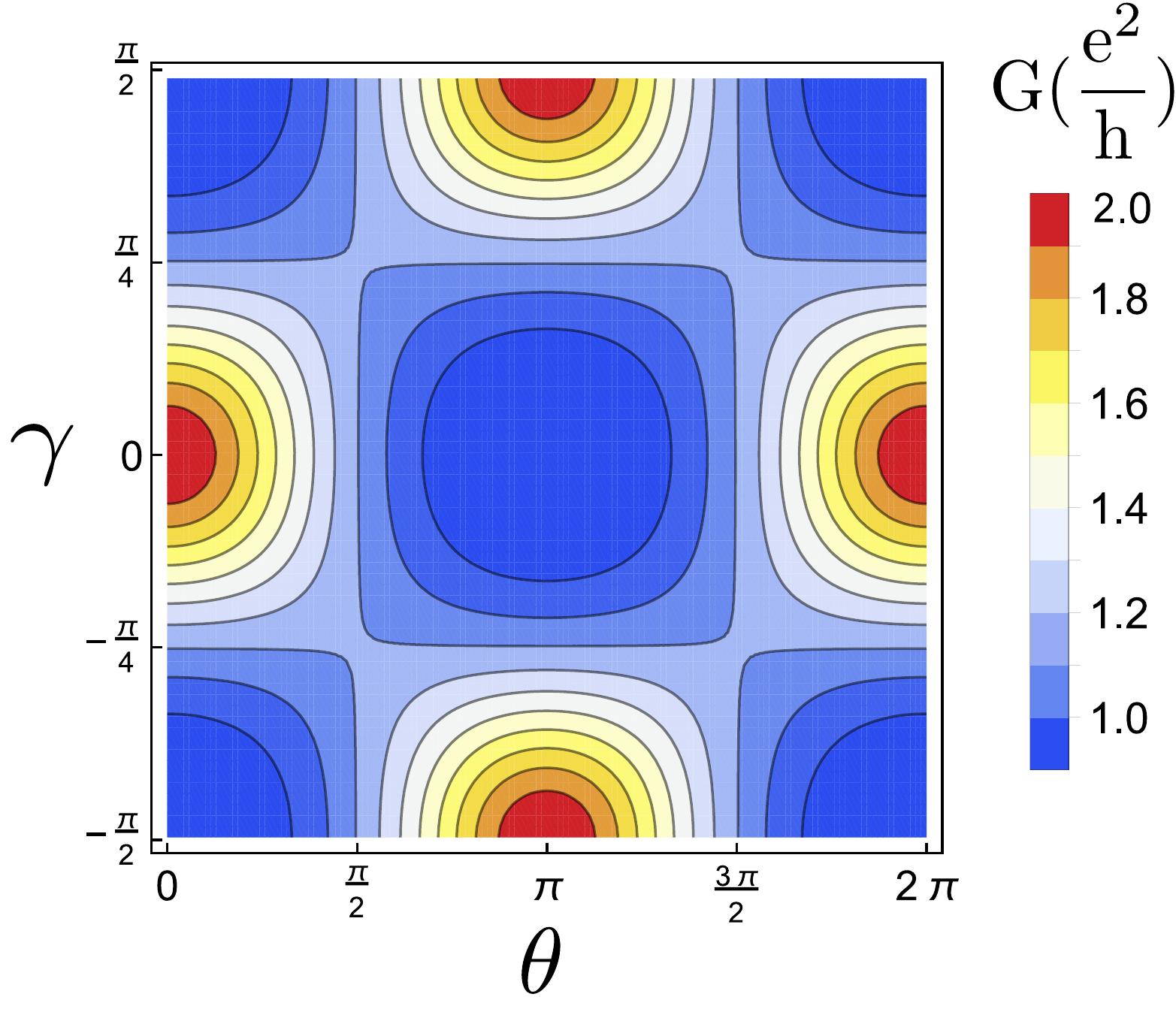}
\caption{Density plot of the unpolarized conductance as a function of the spin tilt $\gamma$ and the magnetization angle $\theta$ of the right ferromagnetic lead.}
\label{dp}
\end{figure}

The transport properties of the system at zero temperature are studied by coupling symmetrically the QSH disk to two semi-infinite ballistic leads. By applying a low bias we calculate the unpolarized conductance using the Landauer formula \citep{Datta},
\begin{equation}
G=\frac{e^2}{h}\sum_{\sigma,\sigma'}T_{\sigma\sigma'}
\label{Landauer1}
\end{equation}
where $e$ is the electron charge, $h$ is the Planck constant and $T_{\sigma\sigma'}$ denotes the transmission probability between incoming $\sigma$ and outgoing $\sigma^{\prime}$ states in the leads.
Following closely the setup of Ref. \onlinecite{Mac2010}, we consider ferromagnetic leads with in-plane magnetization in order to inject polarized spins in the left lead and detect spins polarized along the polar angle $\hat{\theta}$ in the right lead. A schematic picture of the setup is shown in Fig. \ref{setup}. The Hamiltonian of the ferromagnetic leads, $\mathcal{H}_{FM}=\frac{p^2}{2m}\sigma_0+\mathbf{M}(\theta)\cdot\pmb{\sigma}$, contains a Zeeman-splitting term proportional to the magnetization vector $\mathbf{M}(\theta)=M(\cos\theta,\, \sin\theta,\, 0) $ and a vector of Pauli matrices $\pmb{\sigma}$. Eigenstates in the left lead have the form $\ket{ x}_\pm=\frac{e^{ik x}}{\sqrt{2}}(1,\pm 1)^T$, while in the right lead $\ket{\theta}_\pm=\frac{e^{ik x}}{\sqrt{2}}(e^{-i\theta},\pm1)^T$. The QSH region is described by Eq. \ref{H1D}, where in the open geometry setup the quantum number $m$ labeling the eigenstates in Eqs. \ref{eig1} and \ref{eig2} will no longer be quantized. 
\begin{figure}[tbp]
\includegraphics[width=\columnwidth]{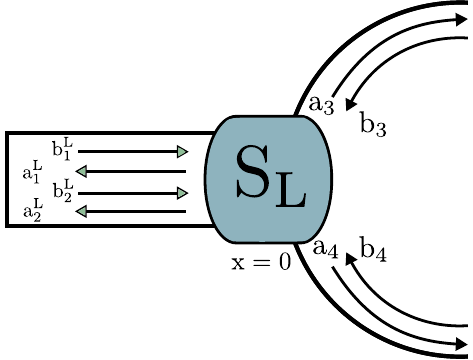}
\caption{Schematic representation of the incoming  and outgoing modes in the left junction of the system.}
\label{Sm}
\end{figure}
The scattering matrix of the system can be obtained from the knowledge of the scattering matrices at the QSH-injector interface $S_L$, and at the QSH-detector interface $S_R$. To calculate them we first notice that if both the spin majority and spin minority bands are occupied
there are two right-moving states and two left-moving states in the ferromagnetic leads. The same number of propagating states are 
found in the QSH disk: each arm of the disk has one clockwise mover and one counterclockwise mover with opposite spin projections.
(Fig.~\ref{Sm}). Hence $S_L$ and $S_R$ are $4 \times 4$ matrices whose elements can be calculated imposing current conservation for each scattering state at the interface. This condition leads to 16 equations for the 16 elements of $S_{L,R}$. For example the scattering ansatz for a right-moving state in the left junction can then be written as,

\begin{equation*}
\begin{cases}
\phi^\lambda_{FM,\mathcal{R}}(x)=\frac{\chi_{FM,\mathcal{R}}^{\lambda}}{\sqrt{|v_{FM}^\lambda|}}e^{ik_\lambda x}+\sum_{\lambda'}r_{\lambda,\lambda'}\frac{\chi_{FM,\mathcal{L}}^{\lambda'}}{\sqrt{|v_{FM}^{\lambda'}|}}e^{-ik_{\lambda' }x} 
\\
\phi^\lambda_{QSH,\mathcal{R}}(x)=\sum_{\lambda'}t_{\lambda,\lambda'}\frac{\chi_{QSH,\mathcal{R}}^{\lambda'}}{\sqrt{|v_{QSH}^{\lambda'}|}}e^{ik_{\lambda'} x} 

\end{cases}
\label{scatteransatz}
\end{equation*}
where $\lambda=\pm$ labels the two possible modes, which in general will have different spinorial parts $\chi^\lambda$ (see Eqs.~\ref{eig1} and \ref{eig2}) and velocities  $v^\lambda$. The indices $\mathcal{R},\mathcal{L}$ discriminate between right-moving and left-moving states.
Since the lead and the disk are parametrized by two different coordinate systems we choose to label the eigenstates with their wave number $k$, which can be simply written as the ratio between the angular momentum $m$ and the disk radius $r_0$. The coefficients $r_{\lambda,\lambda'}$ and $t_{\lambda,\lambda'}$ are the probability amplitudes that a state $\lambda$ will be reflected or transmitted in a state $\lambda'$. Each propagating state is normalized to unit flux in order to obtain a unitary S-matrix. A similar ansatz holds for the right junction.\\
\indent Once both S-matrices are calculated, they can be combined to obtain the full scattering matrix of the device,
\begin{equation*}
\begin{pmatrix}
b^L_i \\[1mm]
b^R_i 
\end{pmatrix}=
\begin{pmatrix}
r && t'\\
t && r'
\end{pmatrix}
\begin{pmatrix}
a^L_i \\[1mm]
a^R_i 
\end{pmatrix}
\end{equation*}
where $b^{L,R}_i$ and $a^{L,R}_i$ ($i=1,2$) are respectively the wave amplitudes of  the two outgoing and ingoing states in the left $L$ and right $R$ lead. 
We define $r$ and $t$ as the $2\times 2$ matrices whose elements are the spin-dependent reflection and transmission amplitudes. The transmission coefficients in Eq.~\ref{Landauer1} are just the modulus square of the elements of $t$. The unpolarized conductance from left to right lead can be then simply expressed as\citep{Datta} $G=\frac{e^2}{h}tr(tt^\dag)$.\\
\indent At zero Rashba, $\gamma=0$, and with both ferromagnets aligned we find perfect transmission, 
$T=diag[1,1]$, when the momenta of the states in the QSH disk satisfy the condition $k r_0=\frac{\mathbb{Z}}{2}$, with $\mathbb{Z}$ an integer.
This resonance effect can be understood by noticing that for $m=\frac{\mathbb{Z}}{2}$ the state inside the QSH disk is also an eigenstate of the closed system, as shown in Section~\ref{sec:effH}. In the remainder, we will fix the Fermi energy of the system to fulfill the resonance condition in the disk. We emphasize that for a sufficiently large disk radius the in gap eigenstates of Eq.~\ref{Eigenvalues1D} are close enough that resonance is achieved for almost any value of the Fermi energy. Calculating $G$ under this assumption yields the density plot in Fig. \ref{dp}, 
where the unpolarized conductance is modulated as a function of the Rashba strength $\gamma$ and the relative magnetization $\theta$ between the two leads. Here the Fermi energy has been fixed close to the bottom of the upper
band in the leads, in order to have a large modulation of the unpolarized conductance. 
If we were to raise the Fermi energy in the leads such that the wave numbers of the two spin-polarized injected electrons were comparable, then the modulation of the conductance would be largely suppressed, and the ballistic conductance, independent of the Rashba strength and the relative magnetization between injector and detector, would be quantized to $2 e^2 /h$.\\
\indent The density plot of the unpolarized conductance  shows that when the ferromagnetic leads have opposite magnetizations ($\theta=\pi$), the maximum conductance is reached as the Rashba field is strong enough
to completely flip the spin of the incoming electrons, that is for $\gamma\rightarrow \frac{\pi}{2}$.
Hence for sufficiently large Rashba couplings the electron spin is reversed: this effect is due to the phase accumulated after half a turn by the two eigenstates in the QSH. From the elements of the $t$ matrix we can also compute the contribution to the conductance of an injected spin polarized current. This polarized conductance turns out to be half
the value of the unpolarized one, regardless of the sign of the spin of the injected carrier.
This suggests that at the resonance condition the conductance modulation is dominated by the spin texture of the helical edge states inside the QSH: at resonance the injected electrons always enter the QSH, but they can only transfer to the right lead if their spin projections can match the ones of the detector. By making use of these observations we can now calculate analytically the spin-polarized conductance.

\subsection{Spin-Polarized Conductance} 
\label{sec:Gan}
Having established that at resonance carriers with opposite spin polarization contribute equally to the conductance, we can restrict ourselves to investigate the modulation of the spin polarized conductance, 
and assume that only the lowest Zeeman band is occupied. Since at resonance injected electrons enter the QSH unimpeded, transmission 
is then determined by the spin projection of electrons upon exiting the disk at $\phi=\pi$. Hence, the conductance modulation is controlled solely by the spin structure of the helical edge states, greatly simplifying the description of the transmission coefficient and leading to a clear analytic understanding of the physics of the system. \\
\indent To calculate the transmission coefficient $T_{\sigma\sigma'}$ we follow the steps of Ref. \onlinecite{Fru2004}. The spin eigenstates $\ket{x}$ incoming from the left lead propagate coherently in the disk, through the helical edge states, and leave the disk in a mixed spin state $\ket{ \sigma_{out}}=\sum_{\lambda=\pm} \bra{\psi_m^\lambda(0)}\ket{x} \ket{\psi_m^\lambda(\lambda \pi)}$. The transmission coefficient can then be obtained from the overlap between $\ket{\sigma_{out}}$ and the outgoing eigenstate $\ket{\theta}$ in the right lead, $T_{x \theta}=\left | \left\langle\theta|\sigma_{out} \right\rangle\right |^2$. 
The polarized conductance takes the form,
\begin{equation}
G_{x\theta}(\gamma, \,\theta)=\frac{e^2}{2h}(1+\cos2\gamma\cos\theta).
\label{Ganalytic}
\end{equation}
The above conductance has the same modulation pattern of the one found in Fig~\ref{dp}. As expected at strong Rashba SOC, $\gamma\rightarrow\frac{\pi}{2}$, the incoming electron spin is completely reversed.\\
\indent To calculate the unpolarized conductance one must sum over all possible spin polarizations in the left and right leads. The calculation yields a conductance $G=\frac{2 e^2}{h}$, which is simply the inverse contact resistance of a single mode conductor. Since the wave numbers of the injected and detected electrons do not play a role in this calculation, this result follows only from the helical nature of the two propagating edge modes: the two modes are orthogonal and cannot interfere with each other. Contrary to a conventional 1D semiconductor ring, an interference pattern is therefore absent. This is due to the fact that in each arm of the semiconductor ring there are double the movers than the ones in the QSH. When the two arms of the ring recombine at $\phi=\pi$, electrons coming from different arms, with the same spin projection, can interfere leading to a modulation of $G$. We point out that, as mentioned earlier, this result is in agreement with the scattering matrix approach analysis assuming the Fermi energy in the leads is such that the wave numbers of the incoming electrons are comparable. We now validate numerically our results by studying a microscopic tight-binding model corresponding to a regularized version of Eq.~\ref{Hbhz}.

\section{2D Tight-Binding Model}
\label{sec:kwant}

\begin{figure}[tbp]
\includegraphics[width=\columnwidth]{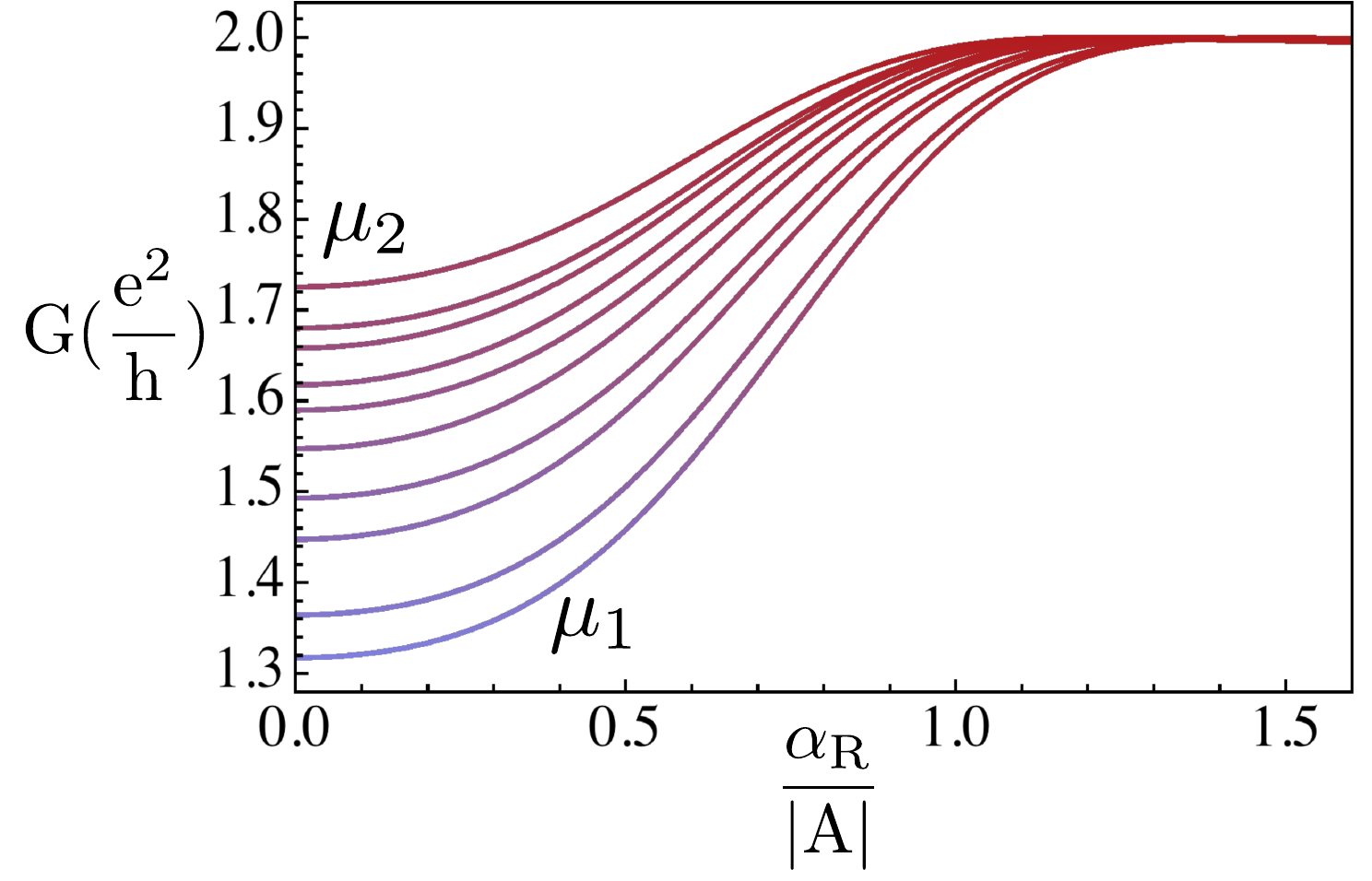}
\caption{Rashba modulated conductance when $\theta=\pi$ in the right lead. In order to inject spin polarized electrons we consider an in-plane magnetic field $H_0$.  The plot shows different modulations of the conductance when varying the chemical potential $(\mu\in[\mu_1,\mu_2]$) in the leads. All energies are normalized to $|A^2/B|$. }
\label{kwantcond}
\end{figure}

To corroborate our findings we perform a numerical calculation of the transport properties of our device. The QSH disk is described using the BHZ tight-binding model on a square lattice,

\begin{align*}
\mathcal{H}^{tb}=&\sum_{ i,\,j}c^\dag_{i,\,j}c_{i,\,j}\hat{V}+
\sum_{ i,\,j}c^\dag_{i+1,\,j}c_{i,\,j}\hat{T}_x \\&+\sum_{ i,\,j}c^\dag_{i,\,j+1}c_{i,\,j}\hat{T}_y+h.c.
\end{align*}

where,
\begin{equation*}
\hat{V}=\mu\, \mathcal{I}_{4 \times 4}+(M-\frac{4B}{a^2})\,\sigma_z\otimes\sigma_0,
\end{equation*}

\begin{equation*}
\hat{T}_x=
\begin{pmatrix}
\frac{B}{a^2} & -\frac{i A}{2a} & -\frac{\alpha_R}{2a} &  0\\[1mm]
-\frac{i A}{2a} & -\frac{B}{a^2} & 0 &0\\[1mm]
\frac{\alpha_R}{2a}& 0& \frac{B}{a^2} & \frac{i A}{2a} \\[1mm]
0& 0 & \frac{i A}{2a} & -\frac{B}{a^2}
\end{pmatrix}
,
\end{equation*}
\begin{equation*}
\hat{T}_y=
\begin{pmatrix}
\frac{B}{a^2} & \frac{A}{2a} & \frac{i\alpha_R}{2a} &  0\\[1mm]
-\frac{ A}{2a} & -\frac{B}{a^2} & 0 &0\\[1mm]
\frac{i\alpha_R}{2a}& 0& \frac{B}{a^2} & \frac{A}{2a} \\[1mm]
0& 0 & -\frac{A}{2a} & -\frac{B}{a^2}
\end{pmatrix}
.
\end{equation*}

Here, $\mu$ is the chemical potential and $a$ is the lattice spacing. The operators $c_{i,j}^\dag$ and $c_{i,j}$ create and annihilate an electron in the lattice site $(i,j)$.\\
The ferromagnetic leads are similarly modelled by $\mathcal{H}^{tb}_{BHZ}$ with $A=0$ and $\alpha_R=0$ in order to decouple $\left | E_1, m_j=\pm\frac{1}{2} \right\rangle$  and $\left | H_1, m_j=\pm\frac{3}{2} \right\rangle$ bands. The ferromagnetic properties of the leads are captured by including a Zeeman splitting term $\hat{V}_Z$ \citep{Rot2010},
\begin{equation*}
\hat{V}_Z=H_0
\begin{pmatrix}
0 & 0 & e^{-i\theta} & 0 \\
0 & 0 & 0 &0 \\
 e^{i\theta} & 0 &0& 0 \\
0 & 0& 0 &0
\end{pmatrix}
\end{equation*}
where $H_0$ is the strength of the magnetic field and $\theta$ is the relative in-plane magnetization angle between the two ferromagnetic leads. The magnetic field only couples to the $\left | E_1, m_j=\pm\frac{1}{2} \right\rangle$ bands, and hence we can tune $H_0$ and the chemical potential in the leads in such a way to only occupy a majority of $\left | E_1, m_j=+\frac{1}{2} \right\rangle$ bands. In this way, even if both spin polarized carriers are injected, we produce effectively a spin polarized current. \\
\indent The conductance of the system has been calculated by using the Kwant code \citep{KWANT}. In Fig.~\ref{kwantcond} we plot the conductance as a function of the Rashba strength, when the two ferromagnetic leads have opposite magnetizations.  For chemical potentials $\mu\sim\mu_1$, the occupied energy bands contain mostly one type of spin polarized carriers. This causes a larger suppression of the conductance for small values of $\alpha_R$, in agreement with the modulation found in Section \ref{sec:smatrix}. For sufficiently large couplings the conductance is again $\frac{2e^2}{h}$ signaling that the incoming electronic spin is being reversed. The spin tilting of the edge states studied in Section \ref{sec:BHZ} reflects directly onto the transport properties of the system. As expected, we observe the same enhancement of the conductance as the one found along the cut at $\theta=\pi$ in the density plot in Fig. \ref{dp}. 
Indeed this numerical modulation of the conductance matches perfectly with the results obtained in Sections \ref{sec:smatrix} and \ref{sec:Gan}.
\section{Conclusions}
\label{sec: conclusions}
In this paper we have shown how a QSH insulator with Rashba SOC can be used to modulate an electric current and manipulate the spin of injected electrons. The setup we propose allows for an all-electric control of the outgoing spin current making it an interesting possibility for a spin field-effect transistor.
It is important to stress that system geometry and size are not particularly relevant as long as the two paths are symmetric and shorter than the phase relaxation length. Moreover, since the QSH edge states are topologically protected, transport is not affected by (weak) nonmagnetic disorder. Electrons will always fully transmit across the QSH and the conductance will only depend on the scattering at the interface between the QSH disk and the leads.
\begin{acknowledgments}
C.O. acknowledges support from a VIDI grant (Project 680-47-543) financed by the Netherlands Organization for Scientific Research (NWO).
\end{acknowledgments}

\end{document}